%% file: acl2023.tex
\pdfoutput=1

\documentclass[11pt]{article}

\usepackage{ACL2023}

\usepackage{times}
\usepackage{latexsym}
\usepackage{subcaption}

\usepackage[T1]{fontenc}

\usepackage[utf8]{inputenc}

\usepackage{microtype}

\usepackage{inconsolata}

\usepackage{hyperref}
\usepackage{multirow}
\usepackage{graphicx}
\usepackage{tabularx}
\usepackage{url}
\usepackage{xcolor}
\usepackage{epsfig}
\usepackage{adjustbox}
\usepackage{amsfonts, amsmath, amssymb}
\usepackage{booktabs} 
\usepackage{comment}
\usepackage{caption, subcaption}
\usepackage{textcomp}
\usepackage{relsize}
\usepackage{stmaryrd}
\usepackage{bbm}
\usepackage{rotating}
\usepackage{helvet}
\usepackage{courier}
\usepackage{cleveref}
\usepackage{xspace}
\usepackage{enumitem}
\usepackage{mdframed}
\usepackage{makecell}
\usepackage{tcolorbox}
\usepackage{nicematrix}
\usepackage{tabu}
\usepackage{colortbl}
\usepackage{xcolor}
\PassOptionsToPackage{table}{xcolor}

\newcolumntype{X}{>{\raggedright\arraybackslash}m{0.9\linewidth}}
\newcolumntype{W}{>{\centering\arraybackslash}m{0.06\linewidth}}
\newcolumntype{L}{>{\arraybackslash}m{0.99\linewidth}}

\input{dfn}

\title{Is Functional Correctness Enough to Evaluate Code Language Models? 
Exploring Diversity of Generated Codes}

\author{Heejae Chon$^{1\ast}$, Seonghyeon Lee$^{2}$\thanks{\ \ Equal contribution}\ , Jinyoung Yeo$^{1}$, Dongha Lee$^{1}$ \\
  $^1$Yonsei University, $^2$Pohang University of Science and Technology \\
  \texttt{\{0914eagle,jinyeo,donalee\}@yonsei.ac.kr}, \texttt{sh0416@postech.ac.kr}}

\begin{document}
\maketitle
\begin{abstract}
\input{000abstract}
\end{abstract}

\section{Introduction}
\label{sec:intro}
\input{010introduction}

\section{Related Work}
\label{sec:relwork}
\input{020relatedwork}

\section{Methods}
\label{sec:eval}
\input{030evaluation}

\section{Experiments}
\label{sec:exp}
\input{040experiments}

\section{Conclusion}
\label{sec:conclusion}
\input{050conclusion}

\section*{Limitations}
\label{sec:limitation}
\input{060limitation}

\section*{Ethical Consideration}
\label{sec:ethical}
\input{070ethical}

\bibliography{anthology,ref}
\bibliographystyle{acl_natbib}

\appendix

\section{Prompt Details}

\subsection{LLM-Eval}
\label{sec:appendixA}
For evaluating the similarity score by LLM-Eval using GPT-4, we adapt the G-Eval framework~\citep{liu2023gpteval}. The descriptions for each score are defined according to the hierarchy we established. Full details in Table\ref{tab:G-eval prmpting}
\input{Appendix/geval_prompt}

\subsection{Planning strategy}
\label{sec:appendixB}
To display prompt strategy's effect on Sim@10, we utilize a part of AlphaCodium 's~\citep{ridnik2024code} framework, self-reflecting, generating possible solutions, and choosing the best solutions. 
In self-reflecting (Table~\ref{tab: Few Shot Prompt}) LLM understands the given description and generates the commentary and explanations in bullet points. 
In Table~\ref{tab: Generating possible solutions} based on the self-reflection and explanation of the previous stage, LLM generates possible solutions to the problem. 
In the final stage(Table~\ref{tab:Choose candidates}, the LLM selects the best solution from the possible solutions by considering the problem and self-reflection. This final solution is the role of guidelines to generate appropriate code for the problem. To ensure consistency, the decoding temperature is set to 0.2 at every step.

\input{Appendix/Reflecting}
\input{Appendix/Generate_possible_solutions}
\input{Appendix/choose_candidates}

\end{document}

%% file: dfn.tex
\newcommand{\gptturbo}{GPT-3.5-turbo\xspace}
\newcommand{\codellama}{CodeLlama\xspace}
\newcommand{\codellamapython}{CodeLlama-P\xspace}
\newcommand{\codellamainstruct}{CodeLlama-I\xspace}
\newcommand{\starcoder}{StarCoder2\xspace}
\newcommand{\starcoderinstruct}{StarCoder2-I\xspace}

\newcommand{\deepseek}{Deepseek-Coder\xspace}
\newcommand{\deepseekinstruct}{Deepseek-Coder-I\xspace}

\newcommand{\apps}{APPS\xspace}

\newcommand{\humaneval}{HumanEval\xspace}

\newcommand{\geval}{LLM-Eval\xspace}

\newcommand{\scc}{SCC\xspace}

\newcommand{\codebert}{CodeBERT\xspace}

\tcbset{
  aibox/.style={
    width=\textwidth,
    top=10pt,
    colback=white,
    colframe=black,
    colbacktitle=black,
    center title,
    fonttitle=\bfseries,
  }
}

\newtcolorbox{AIbox}[2][]{aibox,title=#2,#1}

\tcbset{
  aiboxsmall/.style={
    width=0.98\linewidth,
    top=10pt,
    colback=white,
    colframe=black,
    colbacktitle=black,
    center title,
    fonttitle=\bfseries,
  }
}

\newtcolorbox{AIboxSmall}[2][]{aiboxsmall,title=#2,#1}

%% file: 000abstract.tex
Language models (LMs) have exhibited impressive abilities in generating codes from natural language requirements. 
In this work, we highlight the diversity of code generated by LMs as a critical criterion for evaluating their code generation capabilities, in addition to functional correctness.
Despite its practical implications, there is a lack of studies focused on assessing the diversity of generated code, which overlooks its importance in the development of code LMs.
We propose a systematic approach to evaluate the diversity of generated code, utilizing various metrics for inter-code similarity as well as functional correctness. 
Specifically, we introduce a pairwise code similarity measure that leverages large LMs' capabilities in code understanding and reasoning, demonstrating the highest correlation with human judgment.
We extensively investigate the impact of various factors on the quality of generated code, including model sizes, temperatures, training approaches, prompting strategies, and the difficulty of input problems. 
Our consistent observation of a positive correlation between the test pass score and the inter-code similarity score indicates that current LMs tend to produce functionally correct code with limited diversity.


%% file: 010introduction.tex
Language models (LMs) pre-trained on massive-scale code corpora show remarkable performance in generating codes from natural language requirements. 
Extensive research has explored various methods to enhance the functional correctness of code generated by these models~\citep{ austin2021program, wang2022mconala, chen2021evaluating}.
As a result, recent benchmarks assessing the functional correctness of generated code~\citep{bigcode-evaluation-harness} show that 4 out of 5 generated codes pass all test cases, confirming their accuracy~\citep{guo2024deepseekcoder}. 
This success highlights the potential of LMs to generate code from high-level specifications.

\begin{figure}[t]
    \centering
     \includegraphics[width=\linewidth]{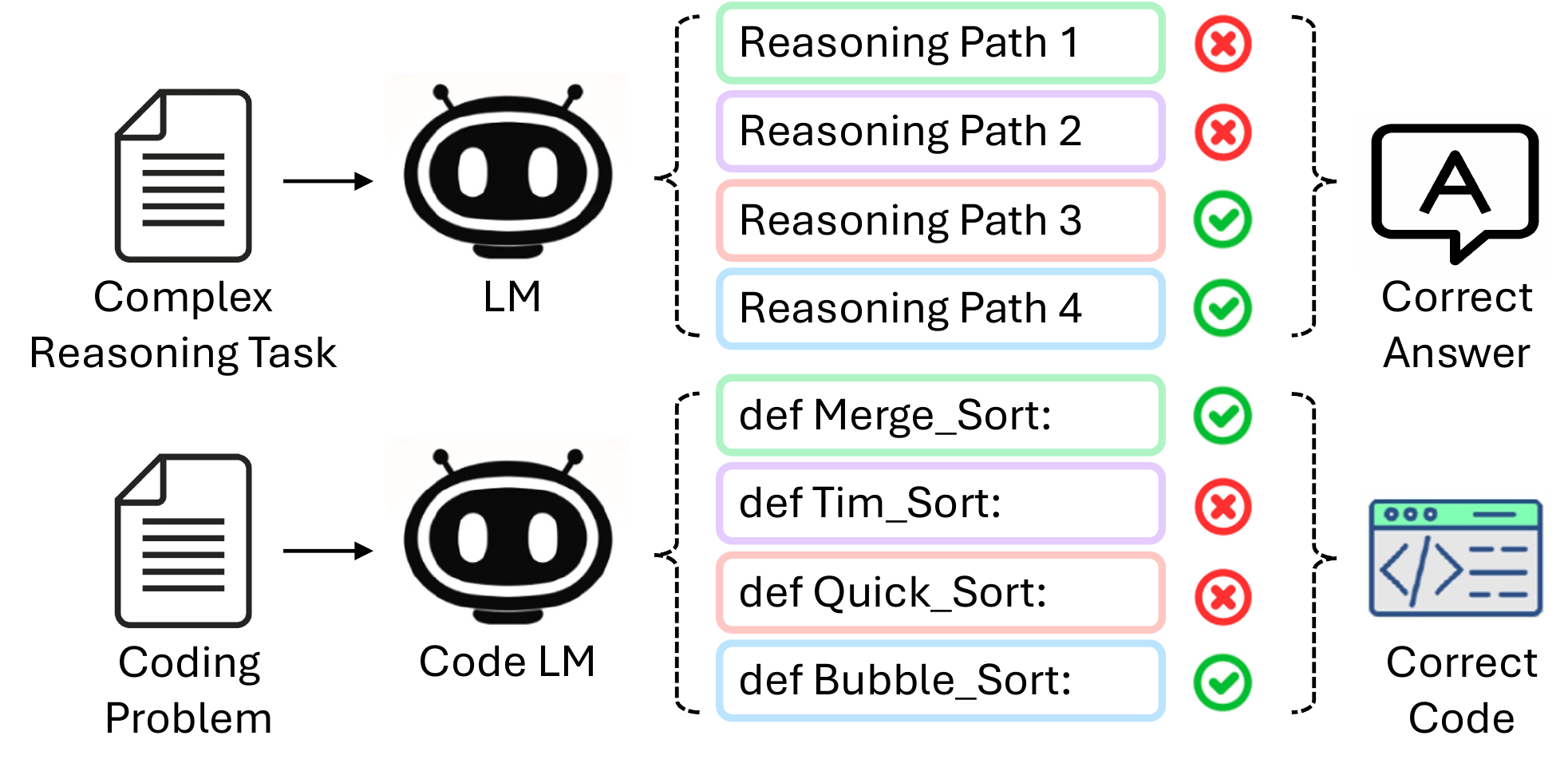}
    \caption{In complex code generation tasks, utilizing diversity encoded in LMs help generate correct outputs.}
    \label{fig:intro} 
\end{figure}

Beyond the significant success of LMs in generating functionally correct codes, the diversity of generated code represents another promising criterion for evaluating the model's code generation abilities, for the following reasons:
First of all, the ability to generate diverse, correct solutions using different approaches for a given problem indicates that the model truly understands the essential requirements and can implement the codes in various ways to fulfill them.
For instance, incorporating diverse reasoning paths during training significantly enhances the code generation capabilities of LMs~\citep{wang2024dolphcoder}.
This aligns with recent studies that emphasize the importance of diverse reasoning paths in enhancing LMs' reasoning capabilities~\citep{zhou2022least, ho2022large}, as illustrated in Figure~\ref{fig:intro}.

Moreover, code LMs are rapidly finding diverse applications these days, and the diversity of the generated code, a fundamental quality metric, is directly influencing software ecosystems.
A report identifies that with the emergence of AI coding assistants, i.e., Copilot,\footnote{{https://github.com/features/copilot}} there has been an increasing trend of repetitive code uploads on GitHub~\citep{gitclear2024};
this trend incurs the degradation of software ecosystems by violating 
a principle of software development, DRY (\textit{Don't Repeat Yourself}).


Nevertheless, there exists scarce studies evaluating the diversity of generated codes, which disregards its importance in developing code LMs.
Most evaluation benchmarks focus only on their functional correctness without quantifying the variety of mechanisms inside their implementations~\citep{chen2021evaluating,hendrycks2021measuring}.
As a result, considerable efforts have been directed towards enhancing code LMs with a lack of emphasis on code diversity~\citep{zheng2023survey}.
This can overlook the potential capability of code LMs or the practical impact of their generated codes.


In this work, we aim to systematically analyze the behavior of competitive code LMs through the lens of the diversity of their generated codes.
To this end, we present a unified framework to evaluate the code generation ability of various LMs, while introducing novel metrics that jointly investigate both code similarity and functional correctness. 
Based on our evaluation framework and comprehensive metrics, we look into the impact of various factors that can affect the quality of generated codes, including model sizes, temperature parameters, training approaches (i.e., instruction-tuned or not), prompting strategies, and difficulties of an input problem (i.e., functional requirements).

Our new metrics primarily focus on 
$K$ sampled codes, similar to the conventional metric for functional correctness (i.e., Pass$@K$). 
Specifically, we introduce two similarity scores, Sim$@K$ and CSim$@K$, which measure inter-code similarity and inter-code similarity conditioned on functional correctness, respectively. 
We also propose a correctness score, DPass$@K$, which quantifies functional correctness by counting the number of correct codes, with DPass further assigning weights based on their diversity.

We incorporate a variety of views for pairwise code similarity into our evaluation framework, which are token-based, embedding-based, and reasoning-based approaches.
In particular, pointing out that existing similarity measures cannot effectively capture high-level code architectures (e.g., algorithm or logic), we introduce a solid approach to utilize LLMs' ability of code understanding and reasoning.
Inspired by the recent success of LLM-based evaluation in text generation~\citep{liu2023gpteval}, we prompt LLMs to thoroughly measure the similarity given a code pair, while considering the hierarchy of code-related concepts;
to this end, we provide detailed instructions as guidance for interpretable scoring.
Through experiments, we validate that this LLM-based measure has a much stronger correlation with human evaluation compared to other measures.



Extensive evaluation of 
two benchmark datasets of varying difficulty levels demonstrate clear and consistent trends in code diversity and functional correctness across various factors. 
Key observations include:
(1) As the difficulty of an input problem increases, the diversity of generated codes also increases across all types of code LMs.
(2) Instruction-tuned models tend to generate more similar codes than their base models, resulting in limited improvement of Pass$@K$ as $K$ increases.
(3) Both decreasing the temperature and adopting enhanced reasoning strategies, such as 
Few-Shot Chain of Thought (CoT) Prompting~\citep{kim2023cot}
, lead to increased inter-code similarity.
For reproducibility, our codes are publicly available.\footnote{{https://github.com/0914eagle/Code-Diversity}}

The main contributions can be summarized as follows:
\begin{itemize}[leftmargin=*,topsep=2pt,itemsep=2pt,parsep=0pt]
    \item This is the first study to thoroughly investigate the diversity of codes generated by code LMs, a topic that has been largely overlooked.
    \item We present a systematic approach to code diversity evaluation, including a reasoning-based pairwise code similarity measure and comprehensive metrics for $K$ sampled codes.
    \item From extensive experiments, we observe that existing code LMs tend to generate functionally correct codes with limited diversity.
\end{itemize}

%% file: 020relatedwork.tex
\subsection{Code Generation with Language Models}
\label{subsec:codelm}
Recently, there have been notable improvements in code generation tasks due to the great advances in language models (LMs).
\citet{austin2021program} and \citet{chen2021evaluating} initially introduced LMs for synthesizing a program and demonstrated their capability by evaluating the functional correctness of generated codes through the unit test.
With this evaluation protocol, researchers have developed the open-source models pre-trained on massive code corpora~\citep{roziere2023code,li2023starcoder,guo2024deepseekcoder} or designed prompt engineering techniques to elicit their code generation capabilities~\citep{kojima2022large,chen2022program,ridnik2024code}, verifying their effectiveness on code generation tasks.

Another research direction focuses on rigorously evaluating the code generation capability by increasing the number of test cases to assess functional correctness~\cite{NEURIPS2023_43e9d647} or by enhancing the requirements to verify robustness in understanding natural language specifications~\citep{wang-etal-2023-recode}.
Additionally, \citet{zheng2024beyond} evaluates the quality of generated codes with various factors to assess the model's ability to meet user demands. 
In this work, we propose a crucial yet often overlooked aspect of code generation capability, i.e., code diversity, by introducing a novel evaluation procedure, which has not been adequately addressed by the research community.

\subsection{Similarity Measure between Codes}
Predicting the semantic relevance between two codes has been a long-standing research topic in code clone detection tasks~\citep{feng-etal-2020-codebert,lu2021codexglue,zakeri2023systematic}.
Especially, \citet{roy2007survey} designed criteria for code similarity measure with varying levels from matching exact codes to consider the functional equivalence between two syntactically different codes.
With these levels, a variety of techniques have been developed to measure code similarity, e.g., token-based~\citep{7886988}, learning-based~\citep{10.1016/j.asoc.2022.109562}, or hybrid techniques~\citep{8816761}.
Another approach relies on LMs' code understanding capability by prompting them to detect clones between two 
generated programs~\cite{dou2023towards}.
Building on this line of research, we integrate these similarity measures into a unified framework (Figure~\ref{figure_2}) and develop an enhanced similarity measure with large LMs while focusing on the criteria from \citet{roy2007survey} to estimate the diversity of generated codes and inspect them in a multifaceted view.

%% file: 030evaluation.tex
In this section, we present a systematic approach to evaluate the code generation ability of LMs in terms of code diversity. 
First, we introduce several pairwise code similarity measures that focus on different aspects and various levels of code-related concepts within a given code pair. 
Using these pairwise similarity measures, we then design a metric that efficiently captures the similarity of $K$ generated codes. 
Finally, we devise a novel metric that jointly considers both code diversity and functional correctness, enabling an investigation into the trade-off between these two aspects.

\begin{figure*}[t]
    \centering
    \includegraphics[width=0.95\linewidth]{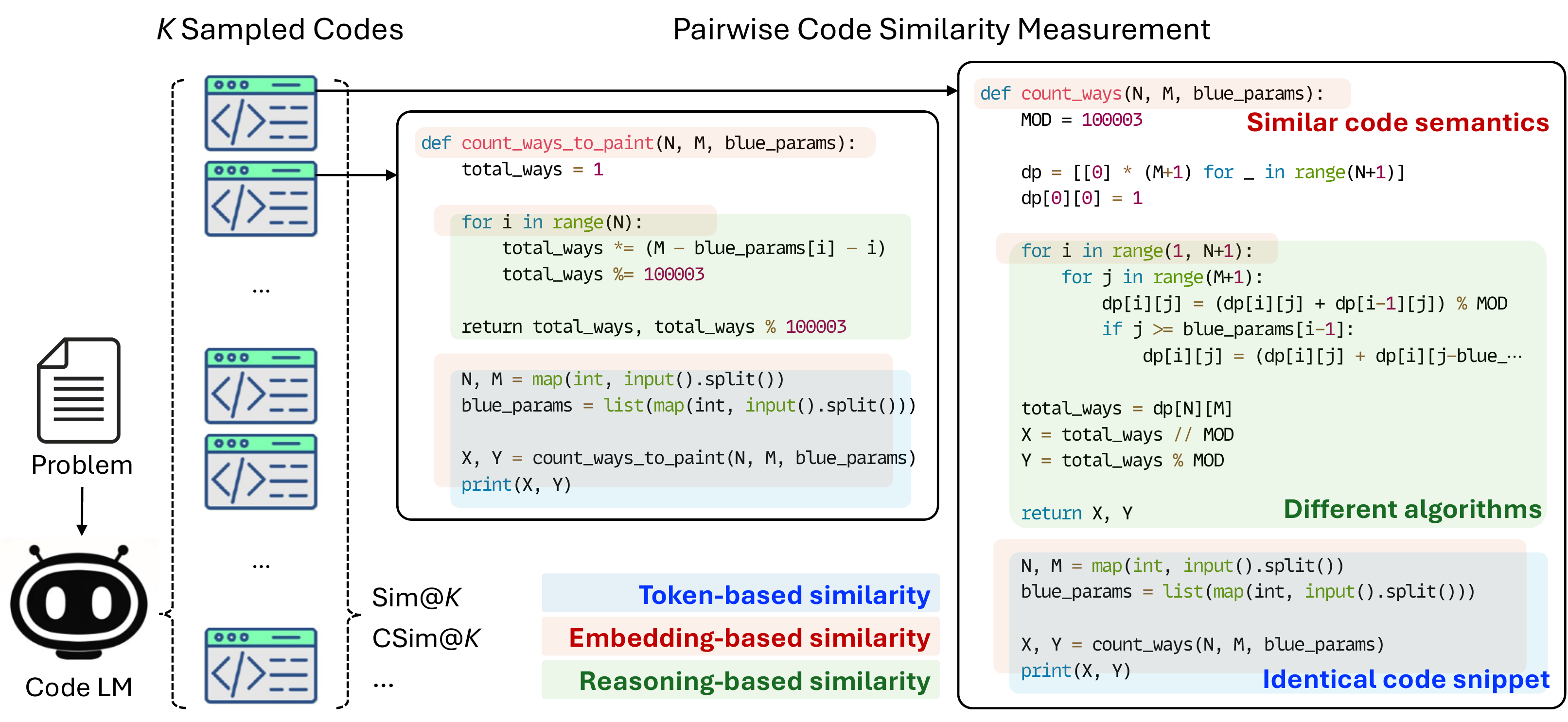}
    \caption{The evaluation pipeline for an LM's code generation ability in terms of inter-code \textit{similarity} (i.e., a negative indicator of \textit{diversity}) while focusing on its $K$ generated codes, termed as Sim@$K$. We adopt three different approaches to pairwise code similarity evaluation, which are based on (1) token-based (2) embedding-based, and (3) reasoning-based similarity measures.}
    \label{figure_2} 
\end{figure*}

\subsection{Pairwise Code Similarity Measurement}
\label{subsec:pairsim}
We present three different approaches for modeling the proximity between two pieces of codes, capturing various levels of code characteristics, such as lexical, syntactical, semantical, and logical similarities.

\subsubsection{Token-based similarity}
\label{subsubsec:tokensim}
The first approach to measuring pairwise code similarity is to use off-the-shelf clone code detectors that mainly rely on lexical similarity between codes.
In this work, we employ a popular open-source detector 
, SourcererCC~\citep{7886988}, which determines if two pieces of code are the same by primarily considering their token-based proximity.
Notably, to effectively handle the presence of comments, which frequently appear within code but do not affect its functionality, SourcererCC automatically detects the comments and removes them before comparing the two codes. 
We use its binary predictions (i.e., 0 for non-copy code, 1 for copy code) with a pre-specified threshold as the measure of token-based similarity.

\subsubsection{Embedding-based similarity}
\label{subsubsec:embsim}

As a pairwise code similarity measure, another approach that captures the semantics within code using pre-trained code encoders is considered. 
Specifically, we compute the contextualized embeddings of codes by employing the bi-directional encoder ~\citep{reimers-2019-sentence-bert}, using their cosine similarity as the measure. 
To acquire embeddings familiar with code, we use an encoder further trained on code search datasets~\citep{husain2019codesearchnet}.\footnote{{https://huggingface.co/flax-sentence-embeddings/st-codesearch-distilroberta-base}} 
Similar to token-based similarity, 
comments are removed manually before evaluating code similarity.

\subsubsection{Reasoning-based similarity}
\label{subsubsec:undersim}
To focus on the core architecture of codes (e.g., algorithm and logic) instead of superficial code texts and their semantics, we propose a novel similarity measure that comprehends and contrasts codes with the help of the reasoning ability of large language models (LLMs).
Motivated by the pioneering work that commands powerful language models to evaluate the quality of generated texts~\citep{liu2023gpteval}, we design a prompt specialized in the coding domain and ask LLMs to rate the similarity between two codes while grounding the detailed guidelines in the prompt.
As the evaluator, we use \texttt{GPT-4}, which effectively follows instructions and demonstrates remarkable performance in code understanding and reasoning tasks, such as code summarization, generation, editing, debugging, and feedback generation~\citep{bubeck2023sparks}.

\begin{figure}[t]
    \centering
    \includegraphics[width=0.8\linewidth]{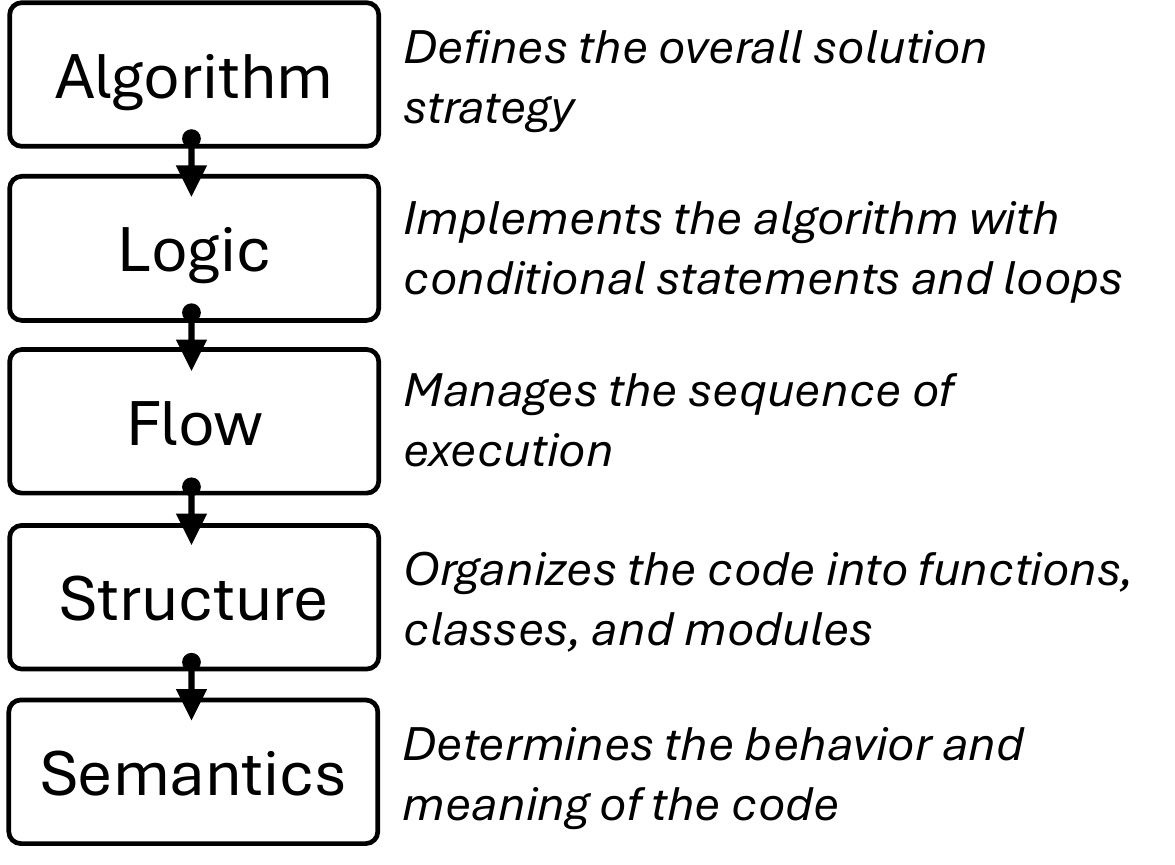}
    \caption{Code-related concepts and their hierarchy from a software engineering perspective.}
    \label{fig:hierarchy} 
\end{figure}

Our prompt is designed to evaluate input code pairs by their similarity on a scale from 1 to 5, emphasizing code understanding and reasoning. 
To provide detailed explanations for each score, which facilitates in-context prompting~\cite{lampinen2022can}, we focus on the hierarchy of code design, clarifying the hierarchical relationship of various code-related concepts with different granularity, such as algorithm, logic, flow, structure, and semantics (Figure~\ref{fig:hierarchy});
this offers clear guidelines, especially for the more ambiguous scores of 2, 3, and 4.
It is worth noting that expertise in programming, i.e., understanding the aspects that make two codes similar to each other, 
is crucial for accurately distinguishing between different levels of similarity.
Key descriptions for each score are presented in Table~\ref{tbl:criteria}, and the full prompt is provided in Appendix~\ref{sec:appendixA}.

\begin{table}[thbp]
\small
    \centering
    \resizebox{1.\linewidth}{!}{%
    \begin{tabular}{cX}
    \toprule
    \textbf{Score} & \textbf{Description} \\
    \midrule
    5 & \textbf{Exact Copy:} Identical in all aspects (code structure, logic, algorithm, comments, whitespaces). \\ \midrule
    4 & \textbf{Minor Modifications:} Trivial changes (variable names, literals, comments) but same structure and logic. \\\midrule
    3 & \textbf{Simple Changes:} Core logic and algorithms preserved with simple control/data flow changes (loop types, order of conditionals). \\\midrule
    2 & \textbf{Substantial Modifications:} Core algorithm maintained with significant changes (extensive comment modifications, rearranged/removed code blocks). \\\midrule
    1 & \textbf{Completely Different:} Entirely different logic and structure (no similarity in code structure, control flow, and data representation).  \\ \bottomrule
    \end{tabular}
    }
    \caption{Summary of pairwise code similarity criteria for the scores (1-5), based on varying levels of code component similarity.}
    \label{tbl:criteria}
\end{table}

\subsection{Similarity Metric for $K$ Sampled Codes}
Based on the pairwise code similarity measures, we define a quantitative metric for evaluating LM's code generation ability in terms of similarity.
Similar to conventional metrics for functional correctness (e.g., Pass@$K$) that focus on the codes generated via $K$ repetitive sampling, our similarity metric named Sim@$K$ mainly assesses the inter-code similarity in the set of $K$ implementations.
Formally, for each problem (or functional requirement) and LM parameters $\theta$, we first obtain the set of $K$ generated codes, $\mathcal{C}_K = \{c_k\sim p_\theta(\cdot)| k=1,\ldots,K\}$,
then we simply compute the average of all pairwise code similarities within the code set.\footnote{To avoid redundant computations, we evaluate each code only once, rather than considering it for all possible pairs. We empirically found that there is no significant difference.} 
Given any pairwise code similarity measure sim($\cdot$, $\cdot$) in Section~\ref{subsec:pairsim}, the overall code similarity score at $K$ sampled results, Sim@$K$, is obtained by 
\begin{equation}
\label{eq:simscore}
\text{Sim}@K = \underset{c_k, c_{k'}\sim\mathcal{C}_K}{\mathbb{E}} \text{sim}(c_{k}, c_{k'}). 
\end{equation}
Note that this inter-code similarity metric serves as a negative indicator of the inter-code diversity metric. 
Therefore, in this paper, we use the terms ``diversity'' and ``similarity'' interchangeably when describing a quality aspect of the generated codes.

\subsection{Joint Metric for $K$ Sampled Codes: Similarity~and Functional Correctness}
To thoroughly evaluate LMs' code generation ability while jointly considering both similarity and functional correctness, we introduce novel metrics to evaluate the $K$ generated codes based on (1) inter-code similarity conditioned on their correctness (denoted by CSim@$K$) and (2) diversity-weighted accuracy (denoted by DPass@$K$).

We first construct the set $\mathcal{C}'_K$ by examining the functional correctness of each code in $\mathcal{C}_K$, i.e., ${\mathcal{C}'_K} = \{c | \mathbb{I}[c\rightarrow \text{Pass}], c\in\mathcal{C}_K\}$.
Then, similar to Sim@$K$, we define CSim@$K$ by the inter-code similarity for the code set $\mathcal{C}'_K$ as follows:
\begin{equation}
\label{eq:csimscore}
    \text{CSim}@K = \underset{c_k, c_{k'}\sim{\mathcal{C}'_K}}{\mathbb{E}}\text{sim}(c_{k}, c_{k'}).
\end{equation}
Note that this score is designed to measure the diversity of the codes conditioned on functional correctness.
Moreover, to examine how well code LMs can generate correct implementations in diverse ways, we define the diversity-weighted correctness metric by
\begin{equation}
\label{eq:acc}
\begin{split}
    \text{DPass}@K &= \text{Pass}@1 \cdot (1-\text{CSim}@K).
\end{split}
\end{equation}
Particularly, the DPass$@K$ metric assigns the weight to the correct codes, proportional to their diversity (i.e., $1-\text{CSim}@K$).
We remark that all these metrics above are calculated for a given problem and then averaged over the dataset.

%% file: 040experiments.tex
\input{041maintable}

In this section, we extensively analyze the implementations generated by the code LMs with the following research questions about code diversity.
\begin{itemize}[leftmargin=*,topsep=2pt,itemsep=2pt,parsep=0pt]
\item \textbf{RQ1}: Can recent code LMs generate sufficiently diverse solutions to specific problems?
\item \textbf{RQ2}: Is there a correlation between the diversity and correctness of the generated codes?
\item \textbf{RQ3}: Do advanced code generation strategies enhance both code diversity and correctness?
\end{itemize}

\subsection{Experimental Setup}
\label{subsec:expset}
\paragraph{Datasets}
We utilize two representative datasets that cover diverse coding problems for code generation tasks:
\apps~\citep{hendrycks2021measuring} and \humaneval~\citep{chen2021evaluating}. 
Both datasets are designed to ask the models to implement Python programs based on the requirements written by natural languages.
In particular, \apps categorize their problems into three difficulty levels, i.e., Introductory, Interview, and Competition, and we report the score separately to identify whether the diversity of generated codes varies depending on their difficulty.
\humaneval consists of basic Python coding problems with the function format where the docstring contains the requirements.
Due to the limited computational resources, we randomly sampled 50 problems for each dataset.

\paragraph{Models}
We include several recent competitive LM families in our experimental setup to differentiate their behavior in terms of code diversity.
\begin{itemize}[leftmargin=*,topsep=2pt,itemsep=2pt,parsep=0pt]
\item \textbf{\codellama}~\citep{rozière2024code} is one of the most prominent code LM families. Especially, we employ a variant, i.e., \codellama-P, that is specifically finetuned for Python. To observe the effect on model size, both 7B and 13B models are adopted in our experiments.
\item \textbf{\starcoder}~\citep{lozhkov2024starcoder} is another open-sourced code LM powered by the big-code organizations. They initially show remarkable performance in code generation in terms of functional correctness.
\item \textbf{\deepseek}~\citep{guo2024deepseekcoder} is a competitive code LM family that outperforms other code LMs in recent code generation benchmark~\citep{bigcode-evaluation-harness}.
\item \textbf{\gptturbo}~\citep{brown2020language} is a well-known proprietary LM powered by OpenAI~\footnote{The version of the model used in this experiment is \texttt{gpt-3.5-turbo-0125}.}. We also include this model as they show outstanding performance in code generation tasks.
\end{itemize}
Whole open-sourced model families have instruction-tuned models and we include them to study how the instruction tuning affects the code generation diversity. 
For each models, we generate ten implementations per problem with temperature scaling with varying temperatures, i.e., 0.2 and 0.8, and nucleus sampling with top 0.95 probability and generate at most 512 tokens per generation.

\subsection{Diversity of Generated Codes (RQ1)}
\label{subsubsec:explm}
Table~\ref{tbl:maintable} reports the Sim@10 scores of various code LMs using four different datasets.
We highlight the key observations from the table as follows:

\paragraph{Comparison across different datasets}
We clearly observe that the similarity of generated code varies significantly depending on the dataset. 
In particular, in \apps dataset, whole models exhibit similar trends based on the difficulty level of the problems. 
As the problems become more difficult, the similarity among the generated codes decreases, indicating that most LMs tend to produce more diverse code as the coding problems become hard. 
As challenging problems lead to uncertain token distributions, LMs produce varied reasoning paths and implementations.
For \humaneval dataset, they show higher similarity score compared to \apps dataset.
We speculate the reason for this experimental results as \humaneval dataset consists of basic coding problems and code LMs are likely to encounter similar form of coding examples during their pre-training stage.

\paragraph{Analysis on similarity measures}
Our results also demonstrate that the three code similarity measures capture diverse aspects between two codes.
For example, \codellama-P 13B produces more similar codes than \starcoder 15B on \apps-Competition with \scc and \codebert measures, but we obtain different results with \geval measure.
This implies that \geval properly captures the latent proximity between the generated codes through reasoning, which is supported by the results in Section~\ref{subsec:correl} as well.


\paragraph{Comparison of instruction-tuned LMs with their base LMs}
Instruction-tuned models exhibit, on average, 26\% higher similarity scores compared to their base models across all datasets.
We observe that this occurs because instruction-tuned models consistently use the same function and variable names, as well as maintain a consistent number of functions.
Notably, \codellama shows the largest score difference, indicating that its ability to generate diverse paths is significantly constrained after instruction tuning.




\begin{figure}[t]
    \centering
    \includegraphics[width=\linewidth]{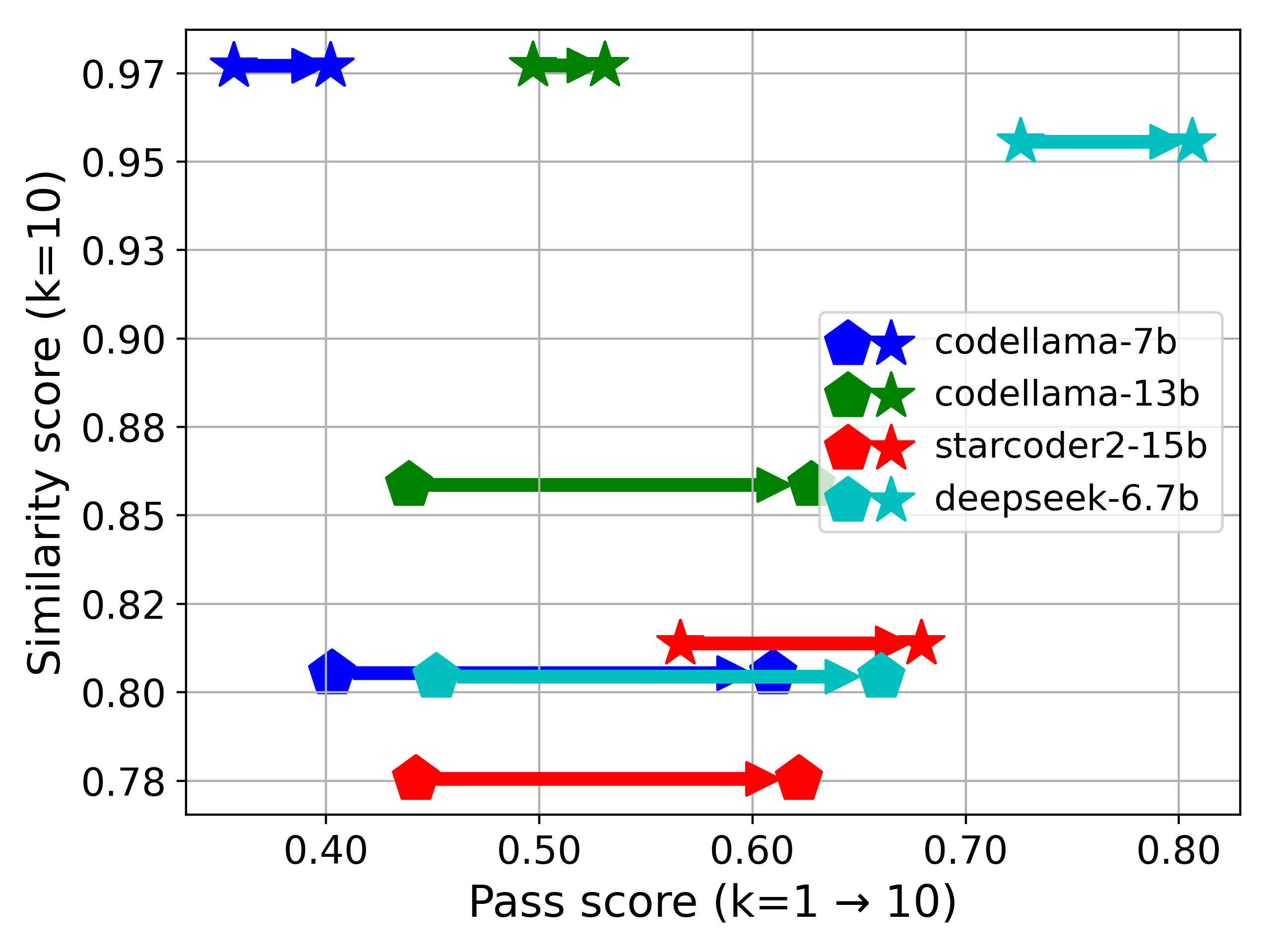}
    \caption{Similarity and functional correctness of generated codes from base models (pentagons) and instruction-tuned models (stars) on \humaneval.}
    \label{fig:pass@1vssim@10}
\end{figure}

\subsection{Relationship between Diversity and Functional Correctness (RQ2)}
\label{subsec:rqtwo}
To further examine the characteristics of generated codes, we study the relationship between the diversity and functional correctness of the generated codes.
To this end, we utilize the Pass$@K$ score, which is widely used to measure functional correctness.
Pass$@K$ score represents the proportion of solved problems where each problem is regarded as solved if, among $K$ candidate implementations, at least one implementation passes the whole test case.

We plot Pass@1, Pass@10, and Sim@10 scores of various code LMs on \humaneval dataset (Figure~\ref{fig:pass@1vssim@10}).
Overall, Pass@1 scores show a noticeable increase as the model fine-tuned to follow instructions.
However, Pass@10 scores for the instruction-tuned models do not increase as Pass@1 scores do.
These results imply that while instruction tuning increases the probability of generating correct codes compared to their base models, it also has the unintended consequence of narrowing the distribution of the implementation space, thereby hindering the model's ability to generate creative or diverse solutions for the given problems.

\begin{figure}[t]
    \centering
    \includegraphics[width=\linewidth]{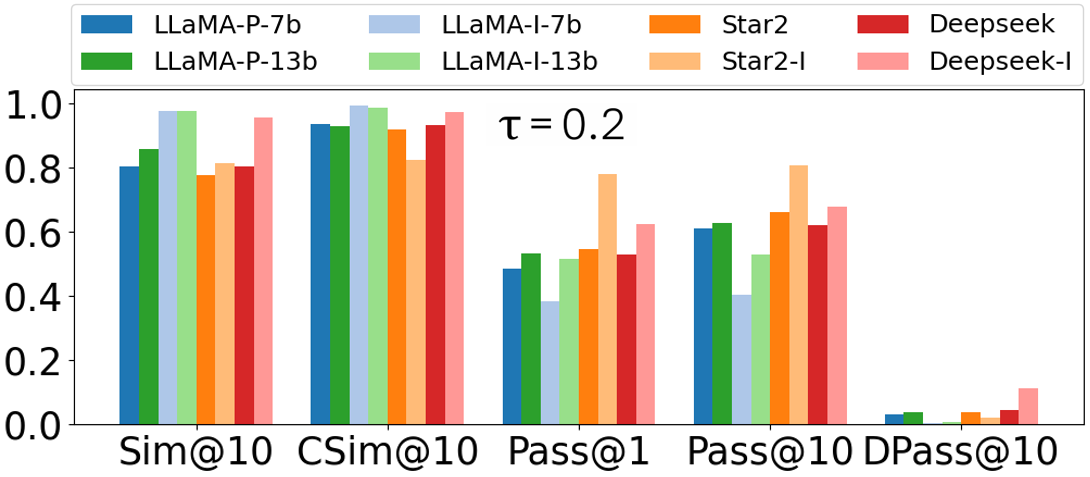}
    \includegraphics[width=\linewidth]{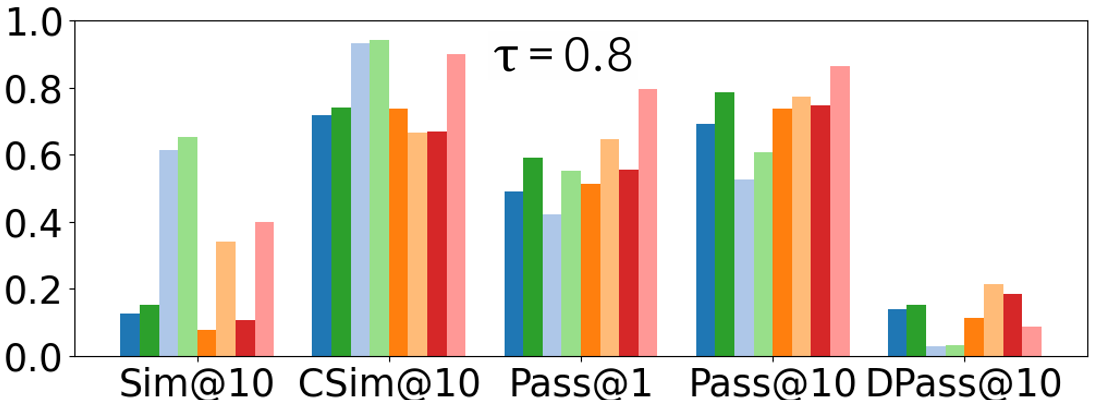}
    \caption{Jointly evaluation of functional correctness and similarity score on the \humaneval dataset.}
    \label{fig:DAcc@10 on the HumanEval dataset} 
\end{figure}

To further consider the diversity among correct implementations, we examine the generated implementations with our proposed metrics in Figure~\ref{fig:DAcc@10 on the HumanEval dataset}.
For all the models, CSim$@10$ scores are generally higher than Sim$@10$ scores, indicating that the correct codes roughly share similar mechanisms.
On the other hand, DPass$@10$ scores are significantly lower, due to the challenge of generating implementations that operate using different correct mechanisms.
Interestingly, increasing the temperature to promote diverse implementations allows \starcoderinstruct to achieve competitive DPass$@10$ scores. 
This indicates that while \starcoderinstruct may be less likely to produce correct code on its own, it can generate a wider range of accurate solutions, suggesting its potential as a generator of diverse and accurate code.

\subsection{Effect of Temperature and Prompting Strategies (RQ3)}
\begin{figure}[t]
    \centering
    \includegraphics[width=\linewidth]{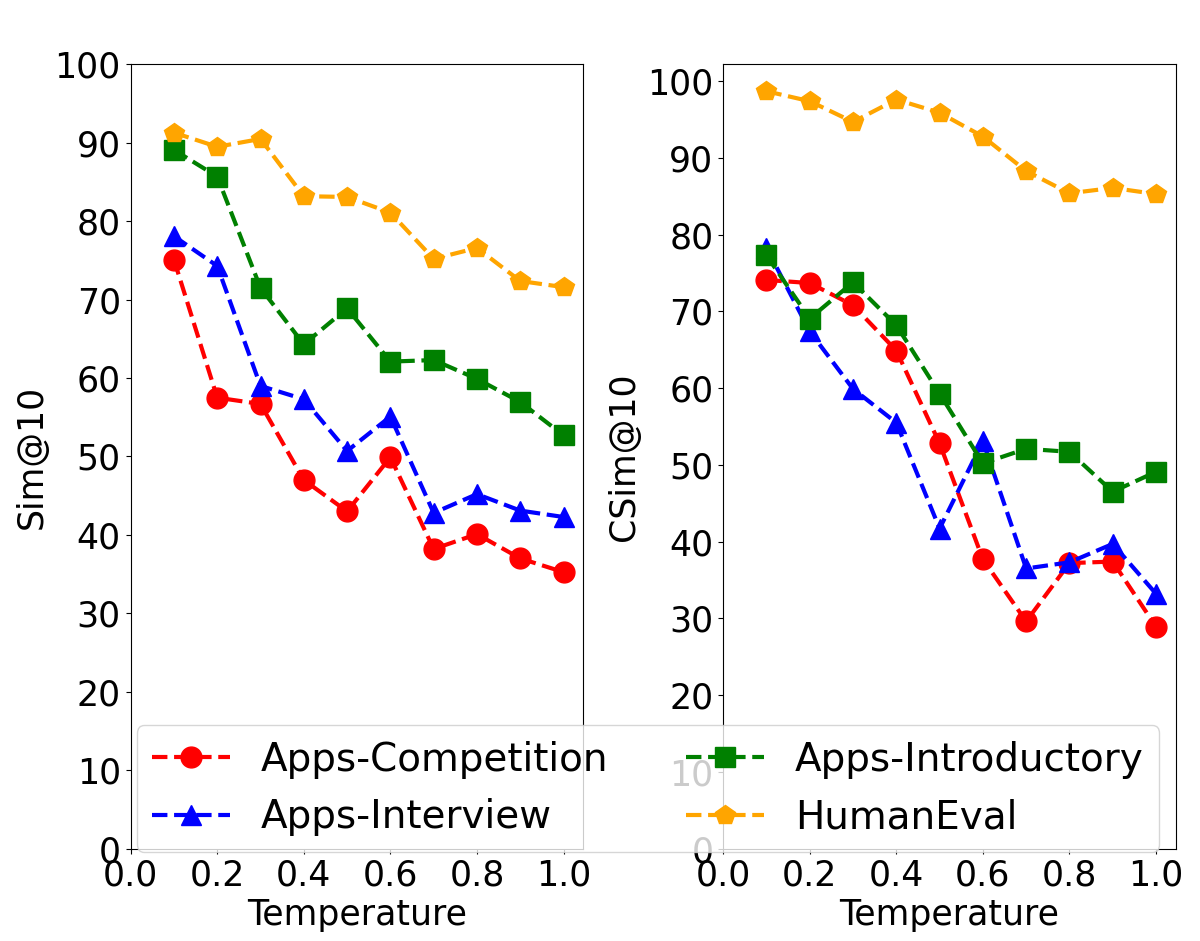}
    \caption{Code similarity changes of \codellama-I 13B with respect to the temperature}
    \label{fig: Temperature effect} 
\end{figure}

We evaluate how the decoding and prompting strategies affect the diversity of generated codes. 
Figure~\ref{fig: Temperature effect} represents Sim$@10$ and CSim$@10$ scores with various temperatures. 
The scores for \humaneval dataset show high CSim$@10$ scores as the problems are easy to solve.
In contrast, the \apps dataset contains more challenging problems, resulting in the generation of more diverse code.
When considering functional correctness, many similar codes are filtered out, causing a tendency for the CSim score to be lower. 
However, all datasets showed a tendency for Sim$@K$ and CSim$@K$ to decrease as the temperature increased, meaning the generated solution became diverse.
This finding is compatible with the previous ones in several recent works~\citep{chen2021evaluating}.

We also examine the effect of few-shot Chain-of-Thought (CoT) prompting~\cite{kojima2022large} and another advanced prompting strategy used in Alphacodium~\cite{ridnik2024code} that enhances the code generation process through its \textit{Planning} step. 
To construct few-shot examples for CoT prompting, we randomly select one problem from each difficulty level in the Apps dataset, resulting in a total of three examples for prompting. 
The planning process involves the GPT-3.5 turbo model generating an explanation and solution for the problem, and then creating candidate codes based on this. Among these candidate codes, the one most similar to the generated solution is used in the GPT prompt.
In Table~\ref{tbl:Prompt Engineering}, applying these prompt strategies leads to an increase in the Sim@10 score, as the generated codes tend to have a more consistent structure, even if they are not functionally correct.
Note that the planning process decreases Pass@1 compared to the vanilla approach, because the performance of \gptturbo is not enough to generate proper explanations or solutions. The full prompt for planning is in Appendix~\ref{sec:appendixB}

\begin{table}[t]
\resizebox{\linewidth}{!}{
\begin{tabular}{lcccc}
\toprule
\multicolumn{1}{c}{Datasets}   & \multicolumn{2}{c}{\textbf{\apps-Introductory}} & \multicolumn{2}{c}{\textbf{\apps-Interview}} \\ \cmidrule{2-5} 
\multicolumn{1}{c}{Model}      & Pass@1 & Sim@10 & Pass@1  & Sim@10 \\ \midrule
Vanilla    &          49.6        &  62.4         &         24.6       &     56.1     \\
\ \ w/ Planning &        34.9      &      77.5     &         \ \ 7.1               &     66.5       \\
\ \ w/ Few-shot CoT     &        57.4        &     79.8    &           45.2          &        67.8    \\ \midrule
\multicolumn{1}{c}{Datasets}   & \multicolumn{2}{c}{\textbf{\apps-Competition}} & \multicolumn{2}{c}{\textbf{\humaneval}}     \\ \cmidrule{2-5} 
\multicolumn{1}{c}{Model}      & Pass@1   & Sim@10   & Pass@1 & Sim@10 \\ \midrule
Vanilla    &        \ \ \ \ 5.8              &     49.3     &      69.9   &    81.3    \\
\ \ w/ Planning &          \ \ \ \ 3.5              &      60.5      &   61.2      &      82.5        \\
\ \ w/ Few-shot CoT     &     \ \    37.3               &     52.8         &   82.4    &     81.5         \\ \bottomrule
\end{tabular}
}
\caption{Effect of prompting strategies to enhance LM's reasoning ability in code generation tasks. GPT-3.5-turbo is used as the target code LM with $\tau=0.2$.}
\label{tbl:Prompt Engineering}
\end{table}

\subsection{Correlation of Human and Automatic Evaluation on Pairwise Code Similarity}
\label{subsec:correl}
We recruit 15 graduate students majoring in computer science and ask them to evaluate the similarity between given code pairs by using their domain knowledge.
No detailed instructions or guidelines are provided, so the scoring criteria rely solely on the judgment of the human evaluators.
Table~\ref{tbl:Expert Eval} summarizes the Pearson correlation between pairwise code similarity measures and human evaluation, based on 100 randomly sampled code pairs spanning a diverse range. 
Our findings indicate that reasoning-based similarity (i.e., \geval) exhibits the strongest correlation with human evaluators' assessments, with its correlation coefficient up to 25\% higher than others. 
In contrast, embedding-based similarity (i.e., \codebert) exhibits no significant correlation with human evaluation (approximately 50\%).
These results strongly suggest that the proposed detailed instructions for LLM prompting can serve as effective guidelines for evaluating pairwise code similarity with a deep understanding of the coding domain.

\begin{table}[t]
\small
\centering
\begin{tabular}{lccc}
\toprule
Measures & \scc &  \codebert & \geval  \\ \midrule
\scc &   -       &    41.4      &    57.9         \\
\codebert     &  41.4  &  -  &         49.5                \\
\geval   &   57.9   & 49.5 & -     \\ \midrule
Human & 54.5 & 51.8 &\textbf{76.8}           \\     \bottomrule
\end{tabular}
\caption{
The Pearson correlation between pairwise code similarity scores from human evaluators and the scores from \scc, \codebert, and \geval. \geval exhibits the highest correlation with human annotators.
}
\label{tbl:Expert Eval}
\end{table}





%% file: 041maintable.tex
\begin{table*}[h]
\centering
\small
\begin{tabular}{lcccccccccccc}
\toprule
\textbf{Datasets} & \multicolumn{6}{c}{\textbf{\apps-Introductory}} & \multicolumn{6}{c}{\textbf{\apps-Interview}}  \\ 
\cmidrule(lr){2-7} \cmidrule(lr){8-13}
Similarity measures & \multicolumn{2}{c}{\scc}  & \multicolumn{2}{c}{\codebert} & \multicolumn{2}{c}{\geval} &\multicolumn{2}{c}{\scc} &  \multicolumn{2}{c}{\codebert} & \multicolumn{2}{c}{\geval} \\ 
\cmidrule(lr){2-3} \cmidrule(lr){4-5} \cmidrule(lr){6-7} \cmidrule(lr){8-9} \cmidrule(lr){10-11} \cmidrule(lr){12-13}
Temperature ($\tau$) & 0.2 & 0.8 & 0.2 & 0.8 & 0.2 & 0.8  & 0.2 & 0.8 & 0.2 & 0.8 & 0.2 & 0.8 \\
\midrule
\codellamapython 7B & 14.0 & \ \ 2.0 & 71.9 & 48.7 & 42.5 & 13.8 & \ \ 8.0 & \ \ 2.8 & 71.6  & 52.1 & 32.5 & 10.5 \\
\codellamainstruct 7B & 52.0 & 27.6 & 89.7 & 81.8 & 65.7 & 53.5 & 50.4 & 14.4 & 92.1 & 80.2 & 60.4 & 42.3\\
\codellamapython 13B & 12.8 & \ \ 3.2 & 71.3 & 51.2 & 39.3 & 22.8 & 10.4 & \ \ 2.4 & 80.4 & 46.9 & 29.8 & 12.3\\

\codellamainstruct 13B & 52.0 & \underline{38.8} & 92.7 & 84.6 & 71.5 & \underline{59.8} & 40.0 & 18.4 & 91.5 & 84.3 & 59.6 & 45.3 \\
\codellamapython 34B & 2.0 & \ \ 1.2  & 46.4 & 46.3 & 14.6 & 14.2 & 1.2 & \ \ 0.4  & 38.8 & 50.0 & 8.9 & 8.1\\
\codellamainstruct 34B & 62.8 & \ \ 24.0  & \textbf{94.6} & 83.7 & 71.6 & 44.8 & \textbf{58.0} & \ \ \underline{20.8}  & \textbf{96.5} & \underline{87.2} & \textbf{70.6} & 41.8 \\
\starcoder 15B & 35.6 & \ \ 3.2 & 80.3 & 52.1 & 49.8 & 22.3 & 14.4 & \ \ 4.8 & 81.7  & 54.8 & 44.8 & 16.9\\
\starcoderinstruct 15B &47.6 & 18.4 & 86.7 & 73.8 & 69.2 & 51.2 & 45.2 & \ \ 3.6 & 84.2 & 62.2 & 57.8 & 29.8 \\
\deepseek 6.7B & 24.4 & \ \ 2.4 & 74.9 & 57.6 & 42.3 & 21.7 & 14.4 
 & \ \ 2.4 & 70.2 & 52.7 & 27.8 & \ \ 9.5 \\
\deepseekinstruct 6.7B & 50.0 & 15.6 & 84.9 & 74.3 & 64.5 & 43.5 & 28.4 & 10.8 & 81.7 & 66.6 & 46.5 & 28.8 \\
\gptturbo & \textbf{65.2}  & 37.6  & 94.4 & \underline{89.1} & \textbf{76.8} & 43.2 & 40.4  & 19.6 & 93.9 & 83.6 & 56.8 & \underline{48.1} \\
\midrule

\textbf{Datasets} & \multicolumn{6}{c}{\textbf{\apps-Competition}} & \multicolumn{6}{c}{\textbf{\humaneval}} \\ \cmidrule(lr){2-7} \cmidrule(lr){8-13}
Similarity measures & \multicolumn{2}{c}{\scc} & \multicolumn{2}{c}{\codebert} & \multicolumn{2}{c}{\geval} &\multicolumn{2}{c}{\scc} &  \multicolumn{2}{c}{\codebert} & \multicolumn{2}{c}{\geval} \\ \cmidrule(lr){2-3} \cmidrule(lr){4-5} \cmidrule(lr){6-7} \cmidrule(lr){8-9} \cmidrule(lr){10-11} \cmidrule(lr){12-13}
Temperature ($\tau$) & 0.2 & 0.8 & 0.2 & 0.8 & 0.2 & 0.8 & 0.2 & 0.8 & 0.2 & 0.8 & 0.2 & 0.8 \\ 
\midrule
\codellamapython 7B & \ \ 5.6 & \ \ 1.6 & 71.2 & 43.8 & 28.3 & \ \ 7.2 & 80.5 & 12.7 & 97.1 & 94.0 & 75.0 & 31.8 \\
\codellamainstruct 7B & 28.8 & \ \ 6.8 & 91.0 & 76.4 & 50.1 & 31.5 & \textbf{97.7} & 61.4 & \textbf{99.9} & 98.0 & \textbf{96.0} & 72.5 \\
\codellamapython 13B & \ \ 9.6 & \ \ 1.6 & 79.6 & 45.1 & 21.5 & \ \ 9.1 & 85.8 & 15.2 & 98.7 & 94.3 & 81.8 & 48.9 \\
\codellamainstruct 13B &  30.8 & 12.8 & 90.8 & 83.8 & 56.8 & 32.1 & \textbf{97.7} & \underline{65.4} & 99.4 & \underline{98.1} & 93.8 & \underline{82.3} \\
\codellamapython 34B & 12.0 & \ \ 0.8  & 70.8 & 44.6 & 27.8 & 6.5 & 47.2 & \ \ 9.2  & 95.3 &  81.7& 74.1 & 42.8 \\
\codellamainstruct 34B & \textbf{48.4} & \ \ \underline{17.6}  & \textbf{93.3} & \underline{85.1} & \textbf{62.5} & 33.5 & 81.6 & \ \ 61.2 & 99.0 & \underline{98.1} & 91.8 & 72.6\\
\starcoder 15B & \ \ 6.4 & \ \ 2.0 & 75.6 & 50.2 & 32.6 & 11.5 & 77.5 & \ \ 7.7 & 96.5 & 93.4 & 76.5 & 42.3 \\
\starcoderinstruct 15B & 33.6 & \ \ 2.4 & 80.4 & 55.5 & 48.3 & 25.5 & 81.3& 34.1 & \textbf{99.9}  & 95.5  & 85.3 & 65.8 \\
\deepseek 6.7B & \ \ 1.2 & \ \ 1.2 & 45.2 & 46.5 & \ \ 8.5 & \ \ 6.8 & 80.4 & 10.6 & 97.4 & 93.5 & 79.3 & 49.5 \\
\deepseekinstruct 6.7B & 31.6 & \ \ 7.2 & 81.5 & 64.8 & 41.5 & 23.8 & 95.5 & 34.1 & 99.9 & 96.5 & 84.3 & 63.5 \\
\gptturbo & 31.6 & 12.0 & 91.3 & 83.2 & 49.6 & \underline{40.1} & 70.8 & 61.2 & 97.2 & 95.3 & 81.2 & 77.3 \\
\bottomrule
\end{tabular}
\caption{Similarity of codes (\%) generated by various sizes of code LMs based on three evaluation methods: \scc, CodeBERT, and LLM-Eval. 
For each (dataset, metric) pair, the highest similarity scores are highlighted in bold for those generated with a temperature of 0.2, and underlined for those generated with a temperature of 0.8.}
\label{tbl:maintable}
\end{table*}

%% file: 050conclusion.tex
In this paper, we explore the code generation ability of LMs by simultaneously evaluating both inter-code similarity (i.e., diversity) and functional correctness.
To achieve this, we propose an evaluation framework that integrates diverse pairwise code similarity measures and comprehensive metrics to assess the quality of $K$ sampled codes. 
Specifically, drawing from software engineering principles, we introduce a reasoning-based similarity measure for code pairs, which exhibits the highest correlation with human evaluators compared to other metrics.
Through extensive experiments, we have shown that the diversity of generated code varies depending on factors such as model sizes, temperatures, training methods, prompting strategies, and problem complexities. However, most existing code LMs tend to produce functionally correct codes that are also similar to each other. 
These observations highlight the necessity for further research aimed at enhancing code LMs to generate more diverse and correct code using a variety of approaches.

%% file: 060limitation.tex
Despite our solid evaluation framework and consistent findings, we identify several limitations in our work and list promising research directions for facilitating research about code diversity.
First, our analysis does not cover large-scale code LMs due to insufficient computational resources. 
As evaluating extremely large models such as CodeLlama 70B requires expensive computational resources, devising an efficient evaluation framework could be a promising future work to enable investigating the behavior of large-scale code LMs and provide valuable insights to the model developers.
Additionally, we randomly sample fifty problems from \apps dataset due to the expensive cost of \texttt{GPT-4}. 
Although our human experiment shows that \geval is highly correlated with humans even with small samples, scaling up this evaluation into whole examples could be useful to some extent.
Lastly, we inevitably adopt a handful of code similarity measures in our experiments even though we extensively search the existing literature about code similarity measures. 
Therefore, developing a code similarity measure that captures different aspects between codes could be useful and synergize with our proposed framework.




%% file: 070ethical.tex
In our research, we have exclusively utilized open-source language models (LMs) and datasets. 
This ensures transparency and allows the broader research community to replicate and build upon our work without legal or proprietary constraints. 
Also, we recruited graduate students in computer science to evaluate code similarity, ensuring their involvement was voluntary and informed. 
We do not intend harm or overlook significant ethical issues.

%% file: Appendix/geval_prompt.tex
\begin{table*}[t]
    \small
    \centering
    \begin{tabular}{p{14cm}}
    \toprule
    \textbf{Simiarity measurement prompt of LLM-Eval} \\
    \midrule
\textcolor{teal}{\textbf{[Task Description]}}\\
You will be given two codes written for a same problem.\\
Your task is to rate the similarity between two codes.\\
Please make sure you read and understand these instructions carefully. Please keep these codes open while reviewing, and refer to it as needed.\\
The output (score) should be integer.\\\\
Evaluation Criteria:\\
Plagiarism (1-5) - the collective quality of code similarity.\\
    \\\\

    \textcolor{teal}{\textbf{[Evaluation Steps]}} \\
1. Read the two codes carefully and identify the main flow.\\
2. Compare between two codes. Check the similarity of these two codes.\\
3. Assign a score for plagiarism on a scale of 1 to 5.\\
\\\\

    \textcolor{teal}{\textbf{[Simiarity scores]}} \\
\textbf{5}: Two codes are an exact copy code. \\
It is copied line by line, with the code structure, logic, algorithms, comments, whitespaces, 
and all other elements being identical.\\
\textbf{4}: Minor trivial modifications have been made.\\
Changes may include variable names, literal values, comments, etc., but the overall code structure, 
logic, and algorithms remain the same.\\
\textbf{3}: While the core logic and algorithms are preserved, there are some simple control or data flow changes.\\
For example, the type of loop (for, while, etc.) may have been changed, or the order of conditional statements may have been altered. \\
However, the overall code meaning remain similar.\\
\textbf{2}: The core algorithm is maintained, \\
but it has been substantially modified to the extent that it has low flow and logic similarity. 
Detailed elements such as comments have been extensively modified, and some code blocks may have been rearranged or removed.\\
\textbf{1}: It is an entirely different, \\
where the logic have been completely changed with no similarity. 
Two code differ in all aspects, including code structure, control flow, and data representation methods \\
even if both codes are solving the same problem.
\\\\
Evaluation Form (scores ONLY):\\
- Plagiarism score:

\\\\\bottomrule
    \end{tabular}
    \caption{The similarity measurement prompt for using LLM-Eval.}
    \label{tab:G-eval prmpting}
\end{table*}

%% file: Appendix/Reflecting.tex
\begin{table*}[h]
    \small
    \centering
    \begin{tabular}{p{14cm}}
    \toprule
    \textbf{Generating self reflection and test explanations} \\
    \midrule
\textcolor{teal}{\textbf{[Task Description]}}\\
"""You are given a code problem:\\
\\
\\
problem description:\\
=====\\
\{description\}\\
=====\\
\\
\\
Given the code problem, you have two tasks:\\
1. Reflect on the problem, and describe it in your own words, in bullet points. Pay attention to small details, nuances, notes and examples in the problem description.\\
2. Explain how each provided example input leads to the corresponding output (in total  [actual-number-of-tests]  examples are provided).\\
Read carefully the problem description. Make sure the test explanations are consistent with them, and between themselves.
The explanation must coherently and logically lead from the input to the output. Be as specific as possible.\\

The output must be a YAML object equivalent to type
ProblemReflection, according to the following Pydantic definitions:\\
=====\\
Class InputOutput(BaseModel):\\
    input: str\\
    output: str\\
    explanation: str = Field(description="Short explanation how the test input leads to the test output.")\\
\\
\\
class ProblemReflection(BaseModel):\\
    self-reflection: str = Field(description="Describe the problem in your own words, in bullet points. Address the problem goals, inputs, outputs, rules, constraints, and other relevant details.")\\
    tests-explanations: list[InputOutput] = Field(max-items={{ actual-number-of-tests }}, description="List of explanations for each test case")\\
=====\\
\\
Example YAML output:\\
```yaml\\
self-reflection:\\
- |\\
  ...\\
- |\\
  ...\\
tests-explanations:\\
- input: |\\
    ...\\
  output: |\\
    ..\\
  explanation: |\\
    ...\\
...\\
 ```\\
\\
\\
Answer:\\
```yaml\\
"""

\\\\\bottomrule
    \end{tabular}
    \caption{The self reflection and test explanation prompt for planning strategy}
    \label{tab: Few Shot Prompt}
\end{table*}

%% file: Appendix/Generate_possible_solutions.tex
\begin{table*}[h]
    \small
    \centering
    \begin{tabular}{p{14cm}}
    \toprule
    \textbf{Generating posssible solutions about code problem} \\
    \midrule
\textcolor{teal}{\textbf{[Task Description]}}\\
"""\\
Pay attention to small details and nuances in the problem description.\\
"""\\
user="""You are given a code problem, and a self-reflection on the problem:\\
\\
problem description:\\
=====\\
\{description\}\\
=====\\
\\
\\
self-reflection on the problem:\\
============\\
\{self-reflection\}\\
============\\
\\
\\
\\
Your goal is to come up with possible solutions to the code problem.\\
\\
Guidelines:\\
- Make sure each solution fully addresses the problem goals, constraints, examples, and notes.\\
- Each solution must have reasonable runtime and memory complexity - less than three seconds on a modern computer, given the problem constraints for large inputs.\\
- Double-check the solutions. Each possible solution must be able to generalize to additional test cases, not just the ones provided in the problem description.\\
\\
The output must be a YAML object equivalent to type ProblemSolutions, according to the following Pydantic definitions:\\
======\\
class Solution(BaseModel):\\
    name: str = Field(description="The name of the solution")\\
    content: str = Field(description="A description of the solution")\\
    why-it-works: str = Field(description="Shortly explain why this solution correctly solves the problem. Be specific and detailed regarding the problem rules and goals.")\\
    labels: List[str] = Field(description="A list of labels for the solution. For example (partial list): binary search, dynamic programming, trees, combinatorics, dfs, bfs, graphs, greedy, math, data structures, geometry, number theory, two pointers, simulation, direct approach, probabilities, ...")\\
    complexity: str = Field(description="The complexity of the solution")\\
\\
class ProblemSolutions(BaseModel):\\
    possible-solutions: List[Solution] = Field(max-items=[max-num-of-possible-solutions], description="A list of possible solutions to the problem. Make sure each solution fully addresses the problem rules and goals.")\\
======\\
\\
Example YAML output:\\
```yaml\\
possible-solutions:\\
- name: |\\
    ...\\
  content: |\\
    ...\\
  why-it-works: |\\
    ...\\
  labels:\\
  - ...\\
  - ...\\
  complexity: |\\
    ...\\
 ```\\
\\
Answer:\\
```yaml\\\
"""

\\\\\bottomrule
    \end{tabular}
    \caption{Based on previous output, self-reflection, prompt for generating possible solutions about the problem description and self-reflection.}
    \label{tab: Generating possible solutions}
\end{table*}

%% file: Appendix/choose_candidates.tex
\begin{table*}[h]
    \small
    \centering
    \begin{tabular}{p{14cm}}
    \toprule
    \textbf{Choosing the best solutions among the candidates} \\
    \midrule
\textcolor{teal}{\textbf{[Task Description]}}\\
"""\\
You are given a code problem and a self-reflection on the problem:\\
\\
problem description:\\
\{description\}\\
\\
self-reflection on the problem:\\
\{self-reflection\}\\
\\
Here is a list of {{ s-possible-solutions|length }} possible solutions to the problem:\\
\{s-possible-solutions-str\}\\
\\
Using the inputs above, your goal is to choose the best solution to the code problem.\\
Don't just pick the most efficient solution. The main consideration is that the solution can fully solve the problem in a simple and robust manner.\\
Make sure the chosen solution has a reasonable runtime - less than three seconds on a modern computer, given the problem constraints regarding large inputs.\\
\\
The output must be a YAML object equivalent to type ProblemSolution, according to the following Pydantic definitions:\\
=======\\
class Test(BaseModel):\\
    input: str\\
    output: str\\
\\
class ProblemSolution(BaseModel):\\
    name: str = Field(description="The name of the best solution")
    content: str = Field(description="The content of the best solution")\\
    why: str = Field(description="Shortly explain why is this the best solution")\\
    flow: List[str] = Field(description="Describe of the flow of the solution, in bullet points")\\
    problem-tests: List[Test] = Field("List the input-output examples that are provided in the problem description.")\\
    input-output-examples-flow: List[str] = Field(description="Describe, in bullet points, how the proposed flow will lead to getting the expected output for the provided input examples")\\
=======\\
Example YAML output:\\
```yaml\\
name: |\\
  ...\\
content: |\\
  ...\\
why: |\\
  ...\\
flow:\\
- |\\
  ...\\
- |\\
  ...\\
...\\
problem-tests:\\
- input: |\\
    ...\\
  output: |\\
    ...\\
input-output-examples-flow:\\
- |\\
  ...\\
- |\\
  ...\\
```\\
\\
Each YAML output MUST be after a newline, indented, with block scalar indicator ('|').\\
\\
Answer:\\
```yaml\\
"""

\\\bottomrule
    \end{tabular}
    \caption{Based on previous solutions, prompt for choosing the best  about the problem description and self-reflection.}
    \label{tab:Choose candidates}
\end{table*}